# Equilibrium Contact Angles and Dewetting in Capillaries

Leonid Pekker[a)], FujiFilm Dimatix, Inc. Lebanon, New Hampshire 03766, USA
David Pekker, University of Pittsburgh, Pittsburgh, Pennsylvania, USA
James Myrick, FujiFilm Dimatix, Inc. Lebanon, New Hampshire 03766, USA

[a)]Author to whom correspondence should be addressed: leonid.pekker@fujifilm.com

**Abstract**

In this work, we extend the model of contact angles that we have previously developed for sessile drops on a wetted surface to the case of a meniscus in a capillary. The underlying physics of our model describe the intermolecular forces between the fluid and the surface of the capillary that result in the formation of a thin, non-removable fluid layer that coats the capillary wall. We describe the shape of the meniscus using a Young-Laplace equation and an incompressible, two-phase, CFD calculation, both modified to take into account intermolecular forces using the disjoining pressure model. We find that our numerical solutions of the Young-Laplace equation and equilibrium meniscus shapes obtained by CFD agree well with each other. Furthermore, for capillaries that are sufficiently larger than the thickness of the non-removable film, our numerical solutions agree well with the effective contact angle model that we previously developed for sessile drops. Finally, we observe that it is possible to tune the disjoining pressure model parameters so that the intermolecular forces between the liquid and solid molecules becomes so strong compared to the surface tension that our formula for effective contact angle gives an imaginary solution. We analyze this situation using CFD and find that it corresponds to dewetting, where the bulk liquid detaches from the walls of the capillary leaving behind the non-removable thin liquid film.

## I. Introduction

In this article, we apply the approach for determining the equilibrium contact angle of a droplet on a wetted substrate to the meniscus in a capillary. As in [1], we describe the cohesive liquid interaction by $\gamma$, the surface tension coefficient, and the net effective liquid-solid interaction of a liquid film of thickness $h$ by the disjoining pressure, $\Pi(h)$, which is the net force per unit area of the liquid-solid interface [2-4]. In Section 2, we consider the case of a slab capillary. We show that, the effective meniscus contact angle increases with an increase in the slab capillary gap but saturates at a constant value given by [1] for capillary gaps greater than approximately 20 times the equilibrium thickness of the film. The analytical model of cylindrical capillary is presented in Section 3. We demonstrate that the cross-sectional meniscus profile in the cylindrical capillary at the transition region between the micro- and the macro-scales differs from the meniscus profile in the slab capillary. At the macro scale, both profiles are well described as a circular/cylindrical cap with the equilibrium contact angle [1]. We observe that, at some parameters of disjoining pressure, the intermolecular forces between the liquid and solid molecules becomes so strong compared to the surface tension, that our formula for effective contact angle gives an imaginary solution. This situation corresponds to dewetting of the wall where the bulk of liquid detaches from the wall leaving at the wall a non-removable thin liquid film. We would like to highlight that usually dewetting is associated with the film rapture, where the nucleation and dry zones are formed in the process of dewetting of a solid surface covered by a thin liquid film. In these processes, the apparent contact angle exists, the liquid nano-scale droplets attach to the solid, and the dry zones are the substrate areas between the droplets covered by a non-removable thin liquid film. Such dewetting processes have been studied extensively both experimentally and theoretically (see for example [5-13]). In our paper, we do not consider thin film rupture processes. In Section 4, we investigate the dewetting process in the capillary using Basilisk, an open-source CFD software, with disjoining pressure implemented using force on interface method. This method has been used in [5] to describe the thin film rupture and dewetting. We present the concluding remarks in Section 5.

## II. Meniscus model in slab capillary

We consider a meniscus in a slab capillary (Fig. 1), with the fluid predominantly on the $z < 0$ side and vapor on the $z > 0$ side and the "tip" of the meniscus located at $z = 0$ and $x = d$, where d is half of the slab capillary gap. The system is translationally invariant along the y-axis. We describe the shape of the lower half of the sessile meniscus by its height $h(z)$ above the bottom wall of the capillary (the upper half is obtained by reflecting the lower half with respect to the plane $x = d$). Neglecting gravitation, the equation describing $h(z)$ can be written as:

$$\frac{d}{dz}\left(\gamma \frac{\frac{d^2 h}{\partial z^2}}{\left(1+\left(\frac{dh}{\partial z}\right)^2\right)^{1.5}} + \chi\left\{\left(\frac{h^*}{h}\right)^m - \left(\frac{h^*}{h}\right)^n\right\}\right) = 0. \tag{1}$$

Eq. (1) states that along the surface of the meniscus the sum of the surface tension ($\gamma$-term) and the disjoining pressure ($\chi$-term) of the meniscus is fixed. The disjoining pressure term is associated with intermolecular force model parameters $\chi = A/(h^*)^3$, $h^*$, $m$, and $n$ of Ref. [5], where, $A$ is the Hamaker constant, $h^*$ is the equilibrium thickness of the non-removable thin film in the absence of a meniscus, and $m$ and $n$ parametrize how the disjoining pressure depends on the film thickness. We ensure that the non-removable film is stable by assuming that $m > n$ so that the disjoining pressure is negative when $h < h^*$ and positive when $h > h^*$. This assumption corresponds to Lenard-Jones intermolecular type potential where the molecules repel each other when the distance between them is small and attract each other when the distance between them is large.

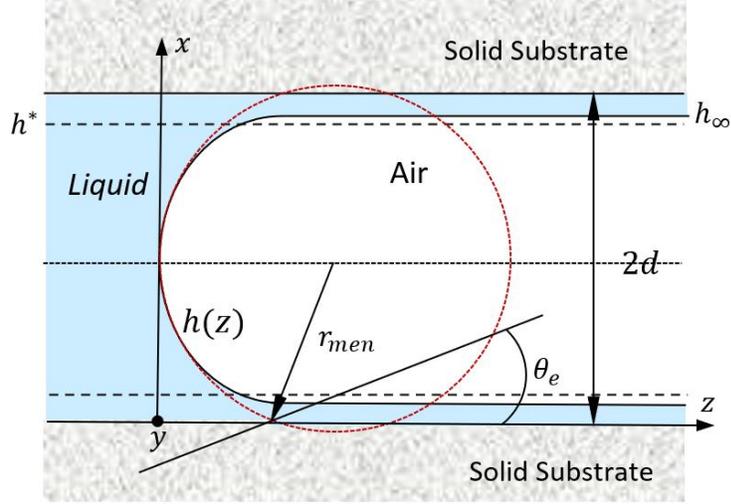

Fig. 1. Schematics of the meniscus model of a slab capillary that supports a non-removable thin liquid film, a longitudinal cross-section of the capillary; $h_\infty$ is the thickness of meniscus far from the center of the meniscus; $h^*$ is the equilibrium thickness of the film; $2d$ is the slab gap; and $r_{men}$ and $\theta_e$ are the radius of the meniscus and the contact angle introduced by Eqs. (17) and Eq. (19) respectively.

Integrating Eq. (1) yields

$$\gamma \frac{\frac{d^2 h}{dz^2}}{\left(1+\left(\frac{dh}{dz}\right)^2\right)^{1.5}} + \chi\left\{\left(\frac{h^*}{h}\right)^m - \left(\frac{h^*}{h}\right)^n\right\} = -P_{men}, \qquad (2)$$

where the constant of integration is the pressure of the fluid in the capillary. We observe that Eq. (2) states that the pressure in the capillary is constant and independent of $z$. On the vapor side, the meniscus gradually becomes a non-removable, fluid film, and therefore the height function asymptote is $h(z) \underset{z\to\infty}{\to} h_\infty$ and $\frac{dh}{dz} \underset{z\to\infty}{\to} 0$ (see Fig. 1). Using this asymptote, we can relate the pressure of the fluid $P_{men}$ to the thin film thickness at infinity $h_\infty$

$$P_{men} = \chi\left\{\left(\frac{h^*}{h_\infty}\right)^n - \left(\frac{h^*}{h_\infty}\right)^m\right\}. \qquad (3)$$

It should be stressed that, unlike in [1], where the pressure in the droplets is always positive and therefore $h_\infty > h^*$, the pressure in the fluid can be both positive and negative for the capillary case. When the meniscus is concave (contact angle is less than $\pi/2$, Fig. 1), the fluid pressure in the capillary is negative and therefore $h_\infty < h^*$, Eq. (3), and when it is convex (contact angle larger than $\pi/2$), the pressure is positive and $h_\infty > h^*$. Let us also introduce an effective meniscus radius, $R_{men}$, as

$$P_{\text{men}} = -\frac{\gamma}{r_{\text{men}}} \qquad (4)$$

We proceed by multiply Eq. (2) by $\frac{dh}{dz}$ and integrating the resulting equation to obtain

$$-\frac{\gamma}{\left(1+\left(\frac{dh}{dz}\right)^2\right)^{0.5}} - \frac{\chi h^*}{m-1}\left(\frac{h^*}{h}\right)^{m-1} + \frac{\chi h^*}{n-1}\left(\frac{h^*}{h}\right)^{n-1} - \chi h \left(\frac{h^*}{h_\infty}\right)^m + \chi h \left(\frac{h^*}{h_\infty}\right)^n = C, \qquad (5)$$

where $C$ is a constant of integration. Application of boundary conditions, $\frac{dh}{dz} \underset{z \to \infty}{\to} 0$ and $h \underset{z \to \infty}{\to} h_\infty$, gives

$$C = -\gamma - \frac{\chi h^* m}{m-1}\left(\frac{h^*}{h_\infty}\right)^{m-1} + \frac{\chi h^* n}{n-1}\left(\frac{h^*}{h_\infty}\right)^{n-1}. \qquad (6)$$

Substituting $C$ from Eq. (6) into Eq. (5) and then solving for $\frac{dh}{dz}$ we obtain an equation describing the shape of the meniscus:

$$\left(\frac{dh}{dz}\right)^2 = \frac{\alpha B_s - (0.5\,\alpha\,B_s)^2}{(1-0.5\,\alpha\,B_s)^2}, \qquad (7)$$

where the subscript "s" indicates slab geometry, and

$$B_s(h) = 2\left(\begin{array}{c}\frac{1}{m-1}\left(\frac{h^*}{h}\right)^{m-1} - \frac{1}{n-1}\left(\frac{h^*}{h}\right)^{n-1} + \frac{h}{h^*}\left(\frac{h^*}{h_\infty}\right)^m - \frac{h}{h^*}\left(\frac{h^*}{h_\infty}\right)^n \\ -\frac{m}{m-1}\left(\frac{h^*}{h_\infty}\right)^{m-1} + \frac{n}{n-1}\left(\frac{h^*}{h_\infty}\right)^{n-1}\end{array}\right), \qquad (8)$$

$$\alpha = \frac{\chi h^*}{\gamma}. \qquad (9)$$

We supplement Eq. (7) by the boundary conditions at the "tip" of the meniscus

$$h(z=0) = d \text{ and } \left(\frac{dh}{dz}\right)_{h=d} = \infty. \qquad (10)$$

Finally, we observe that boundary conditions (10) imply that the denominator of Eq. (7) goes to zero at $z = 0$, which we give us an implicit equation for $h_\infty$

$$1 - \alpha\left(\begin{array}{c}\frac{1}{m-1}\left(\frac{h^*}{d}\right)^{m-1} - \frac{1}{n-1}\left(\frac{h^*}{d}\right)^{n-1} + \frac{d}{h^*}\left(\frac{h^*}{h_\infty}\right)^m - \frac{d}{h^*}\left(\frac{h^*}{h_\infty}\right)^n \\ -\frac{m}{m-1}\left(\frac{h^*}{h_\infty}\right)^{m-1} + \frac{n}{n-1}\left(\frac{h^*}{h_\infty}\right)^{n-1}\end{array}\right) = 0. \qquad (11)$$

We remark that Eq. (7) was obtained in [1], where it was used with different boundary conditions to determine the shape of a sessile droplet on a wetted substrate.

Next, we obtain a formula for the effective radius of curvature of the meniscus, $r_{\text{men}}$ (see Fig. 1), for the case of a large slab capillary gap such that $d \gg h^*$, $|P_{\text{men}}| \ll \chi$ and $\left|\frac{h^*}{h_\infty} - 1\right| \ll 1$. Recasting $h_\infty$ in terms of $h^*$ as

$$h_\infty = h^* (1 - \varepsilon) \tag{12}$$

where $|\varepsilon| \ll 1$ quantifies the deviation of the non-removable film thickness from $h^*$. Substituting Eq. (12) into Eqs. (3) and (4) and using $|\varepsilon| \ll 1$, we obtain

$$P_{\text{men}} = -\chi \varepsilon (m - n), \tag{13}$$

$$r_{\text{men}} = \frac{\gamma}{\chi \varepsilon (m-n)}. \tag{14}$$

Next, we obtain $\varepsilon$ from Eq. (11). Since $d \gg h^*$ and $\varepsilon \ll 1$, in the right-hand side of Eq. (11), we drop the first two terms, use the approximation $(1 - \varepsilon)^l = 1 - \varepsilon l$ in the third and fourth terms, and drop $\varepsilon$ in the fifth and the sixth terms that reduces this equation to

$$1 = \alpha \frac{d}{h^*} (m - n) \varepsilon + \alpha \frac{m-n}{(m-1)(n-1)}. \tag{15}$$

Solving Eq. (15) for $\varepsilon$

$$\varepsilon = \frac{h^*}{\alpha\, d\, (m-n)} \left(1 - \alpha \frac{(m-n)}{(m-1)(n-1)}\right). \tag{16}$$

and then substituting Eq. (16) into Eq. (14), we obtain

$$r_{\text{men}} = \frac{d}{1 - \frac{\alpha (m-1)}{(m-1)(n-1)}}, \qquad P_{\text{men}} = -\frac{\gamma}{d}\left(1 - \alpha \frac{(m-n)}{(m-1)(n-1)}\right). \tag{17}$$

Finally, we determine the effective contact angle $\theta_e$ as tangent of the angle at which the circle $r_{\text{men}}$ crosses the slab wall in Fig. 1,

$$\tan(\theta_e) = \frac{(r_{\text{men}}^2 - d^2)^{1/2}}{d}. \tag{18}$$

Substituting $r_{\text{men}}$ from Eq. (17) into Eq. (18), we obtain

$$\tan(\theta_e) = (\alpha)^{1/2} \frac{\left(\frac{2(m-n)}{(m-1)(n-1)} - \alpha\left(\frac{(m-n)}{(m-1)(n-1)}\right)^2\right)^{1/2}}{1 - \alpha \frac{(m-n)}{(m-1)(n-1)}} \tag{19}$$

As one can see from Eq. (19), $\theta_e$ is independent of the capillary gap and matches the equilibrium contact angle for large droplets obtained in [1].

For the case of a concave meniscus $\left(\theta_e < \frac{\pi}{2}\right)$, as follows from Eqs. (19) and (17), $\alpha < \frac{(m-1)(n-1)}{(m-n)}$ and $r_{\text{men}} > 0$. For the special case of $\theta_e = \frac{\pi}{2}$ and hence $r_{\text{men}} \to \infty$ (Fig. 1), $\alpha = \frac{(m-1)(n-1)}{(m-n)}$ and consequently $\varepsilon = 0$, $h_\infty = h^*$ and $P_{\text{men}} = 0$, Eqs. (16), (12), and (13).

For the case of a convex meniscus case $\left(\theta_e > \frac{\pi}{2}\right)$, as follows from Eq. (19), $\alpha$ is larger than $\frac{(m-1)(n-1)}{(m-n)}$ and, consequently, $\varepsilon$ is negative, Eq. (16). When the meniscus curvature radius is negative as defined in our reference frame, Eq. (17), then $h_\infty > 1$, Eq. (12), and the liquid pressure in the capillary is positive, Eq. (3). We observe that, when $\alpha$ becomes larger than $\frac{2(m-1)(n-1)}{m-n}$, $|r_{\text{men}}|$ becomes smaller than $d$, Eq. (17), and, consequently, $\theta_e$ in Eq. (19) becomes imaginary in value. This situation is unphysical, and corresponds to dewetting, where the liquid column separates from the capillary wall leaving a non-removable thin liquid film with equilibrium thickness, $h^*$. The same situation occurs in the case of a droplet on the substrate. When $\alpha > \frac{2(m-1)(n-1)}{m-n}$, the droplet detaches from the substrate leaving a non-removable thin liquid film. These phenomena are considered in Section IV.

In Fig. 2, we compare the meniscus shapes obtained numerically, with no further approximation beyond those involved in Eq. (1), against the "cylindrical" meniscus model with the meniscus radius $r_{\text{men}}$ obtained from Eq. (17). We numerically integrated Eq. (7), starting from the boundary condition $h(0) = d$, with the value of $h_\infty$ obtained from Eq. (11). Since $(dh/dz)_{z=0} = \infty$, Fig. 1, in the code, in the RHS of Eq. (7), we have used $d - \varepsilon$ with $\varepsilon \ll d$. In constructing our comparisons: we set the inter-molecular force exponents to $m = 9$ and $n = 3$ [6]; we consider two values of the $\alpha$ parameter $\alpha = 0.3$ and $\alpha = 5$ which correspond to $\theta_e = 27.4°$ and $\theta_e = 151.0°$; finally we vary the ratio of $d/h^*$ from 10 to 40. We observe that for both values of $\alpha$ except for the "foot"-like feature at the base of the meniscus where the fluid surface transitions from the cylindrical shape of the meniscus to the flat shape of the non-removable film both models predict very similar meniscus shapes down to $d/h^* \gtrsim 20$ (Fig. 2). On the other hand, for $d/h^* = 10$ we see that the size of the "foot"-like feature becomes comparable to the size of the meniscus and therefore the cylindrical meniscus approximation starts to significantly deviate from the numerical solution.

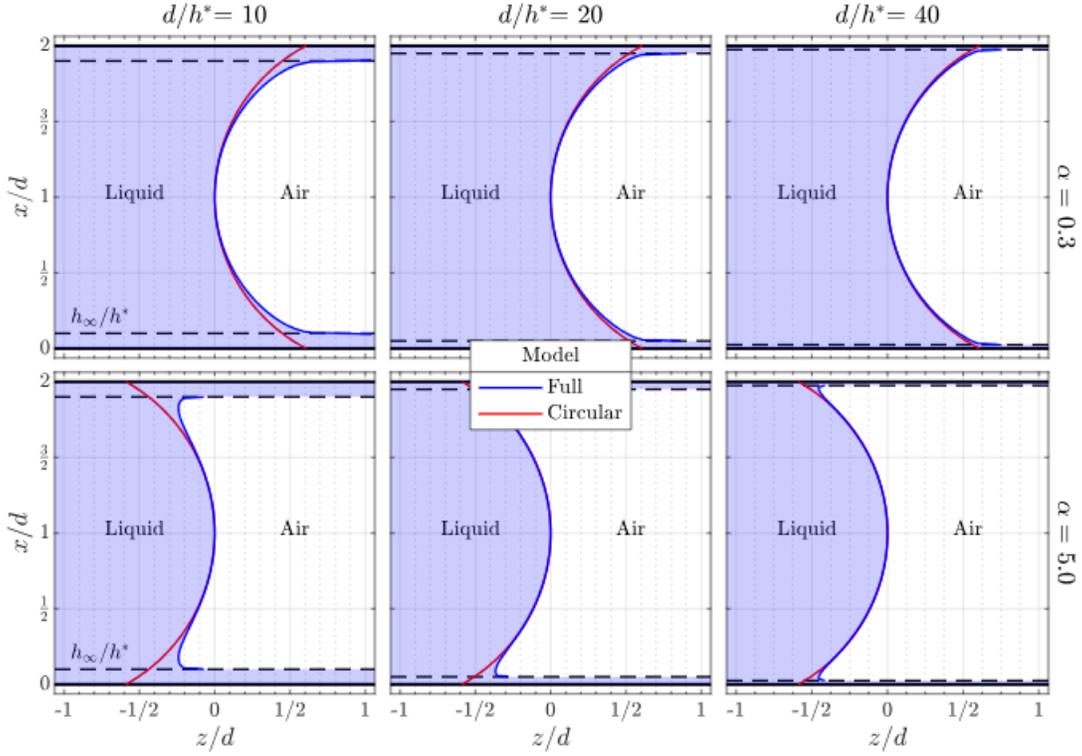

FIG. 2. Meniscus profiles calculated for various values $d/h^*$ and $\alpha_l$ using the full model and a circular meniscus model; $\alpha = 0.3$ corresponds to $\theta_e = 27.4°$ and $\alpha = 5.0$ corresponds to $\theta_e = 151.0°$.

### III. Meniscus model for cylindrical capillary

In this section, we modify the results of the previous section for the case of a cylindrical capillary. Switching the rectilinear coordinate system for the cylindrical one (Fig, 3), Eq. (1) becomes

$$\frac{d}{dz}\left(\gamma \frac{1+\left(\frac{dh}{dz}\right)^2 - h\frac{d^2h}{dz^2}}{h\left(1+\left(\frac{dh}{dz}\right)^2\right)^{1.5}} + \chi\left\{\left(\frac{h^*}{r-h}\right)^m - \left(\frac{h^*}{r-h}\right)^n\right\}\right) = 0, \qquad (20)$$

where r is the radial coordinate and the first term on the left-hand side of Eq. (20) has been modified to describe the surface tension in a cylindrical capillary.

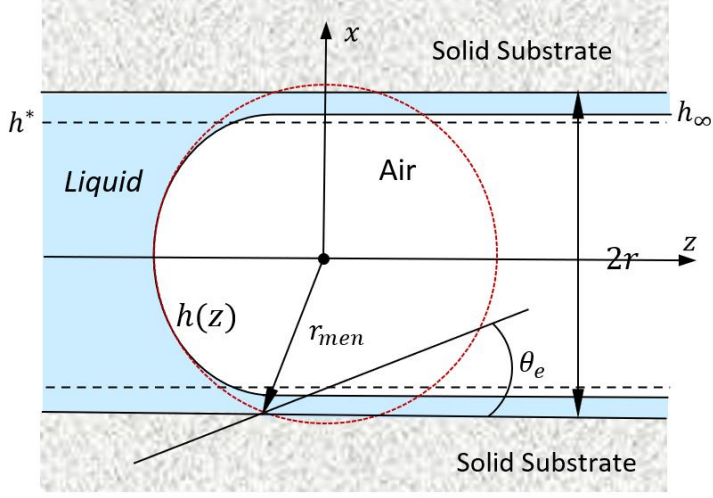

Fig. 3. Schematics of the meniscus model of a cylindrical capillary that supports a non-removable thin liquid film, a longitudinal cross-section of the capillary; $h_\infty$ is the thickness of meniscus far from the center of the meniscus; $h^*$ is the equilibrium thickness of the film; $r$ is the radius of the capillary; and $r_{men}$ and $\theta_e$ are the radius of the meniscus and the contact angle introduced by Eqs. (36) and Eq. (38) respectively.

Integrating Eq. (20) yields

$$\gamma \frac{1+\left(\frac{dh}{\partial z}\right)^2 - h\frac{d^2h}{dz^2}}{h\left(1+\left(\frac{dh}{dz}\right)^2\right)^{1.5}} + \chi\left\{\left(\frac{h^*}{r-h}\right)^m - \left(\frac{h^*}{r-h}\right)^n\right\} = -P_{men} \tag{21}$$

where the constant of integration,

$$P_{men} = -\gamma\frac{1}{r-h_\infty} - \chi\left\{\left(\frac{h^*}{h_\infty}\right)^m - \left(\frac{h^*}{h_\infty}\right)^n\right\} \tag{22}$$

is the liquid pressure in the cylindrical capillary. Next, we introduce an effective meniscus radius $r_{men}$ (Fig. 3) as

$$P_{men} = -\frac{2\gamma}{r_{men}} \tag{23}$$

In case of concave meniscus (Fig. 3), $r_{men} > 0$, and, in convex, $r_{men} < 0$.

Follow the steps describing in Section II, we proceed by multiplying Eq. (21) by $h\frac{dh}{dz}$ and integrating the resulting equation to obtain (see Appendix A)

$$\left(\frac{dh}{dz}\right)^2 = \frac{1-B_c^2}{B_c^2}, \tag{24}$$

$$B_c(h) = \frac{1}{2}\left(\frac{h^*}{r-h_\infty} + \alpha\left\{\left(\frac{h^*}{h_\infty}\right)^m - \left(\frac{h^*}{h_\infty}\right)^n\right\}\right)\frac{h}{h^*} +$$

$$\alpha \frac{h^*}{h} \begin{pmatrix} -\frac{r}{h^*} \frac{1}{m-1} \left(\frac{h^*}{r-h}\right)^{m-1} + \frac{r}{h^*} \frac{1}{n-1} \left(\frac{h^*}{r-h}\right)^{n-1} + \frac{1}{m-2} \left(\frac{h^*}{r-h}\right)^{m-2} - \frac{1}{n-2} \left(\frac{h^*}{r-h}\right)^{n-2} + \\ \frac{1}{2\alpha} \frac{r-h_\infty}{h^*} - \frac{1}{2} \left\{ \left(\frac{h^*}{h_\infty}\right)^m - \left(\frac{h^*}{h_\infty}\right)^n \right\} \left(\frac{r-h_\infty}{h^*}\right)^2 + \\ \frac{r}{h^*} \frac{1}{m-1} \left(\frac{h^*}{h_\infty}\right)^{m-1} - \frac{r}{h^*} \frac{1}{n-1} \left(\frac{h^*}{h_\infty}\right)^{n-1} - \frac{1}{m-2} \left(\frac{h^*}{h_\infty}\right)^{m-2} + \frac{1}{n-2} \left(\frac{h^*}{h_\infty}\right)^{n-2} \end{pmatrix}, \quad (25)$$

where index "c" corresponds to cylindrical geometry and $\alpha$ is given by Eq. (9).

Applying boundary conditions $\frac{dh}{dz} \underset{z \to 0}{\to} \infty$ and $h \underset{z \to 0}{\to} 0$ (Fig. 3) to Eq. (25) we obtain an equation for $h_\infty$ by equalizing the expression in the second brackets in the right-hand side of Eq. (25) to zero at $h = 0$,

$$\frac{1}{(m-1)(m-2)} \left(\frac{h^*}{r}\right)^{m-2} - \frac{1}{(n-1)(n-2)} \left(\frac{h^*}{r}\right)^{n-2} + \frac{1}{2\alpha} \frac{r-h_\infty}{h^*} - \frac{1}{2} \left\{ \left(\frac{h^*}{h_\infty}\right)^m - \left(\frac{h^*}{h_\infty}\right)^n \right\} \left(\frac{r-h_\infty}{h^*}\right)^2 +$$

$$\frac{r}{h^*} \frac{1}{m-1} \left(\frac{h^*}{h_\infty}\right)^{m-1} - \frac{r}{h^*} \frac{1}{n-1} \left(\frac{h^*}{h_\infty}\right)^{n-1} - \frac{1}{m-2} \left(\frac{h^*}{h_\infty}\right)^{m-2} + \frac{1}{n-2} \left(\frac{h^*}{h_\infty}\right)^{n-2} = 0. \quad (26)$$

Noting that at the center of the meniscus where $h \to 0$ and $|dh/dz| \to \infty$, the second term in the right-hand side of Eq. (25) is numerically ill-defined; we name this term as A-term. However, as shown in Appendix B, this term does not have issue while $h \to 0$, we just must use Eq. (27) for this term when $h \to 0$,

$$A = \frac{\alpha}{2} \left( \left(\frac{h^*}{r}\right)^n - \left(\frac{h^*}{r}\right)^m \right) \frac{h}{h^*} \quad (27)$$

Now let us obtain a formula for $r_{men}$ depicted in Fig. 3 for the case of a large capillary radii in which $r \gg h^*$, that corresponds to $|P_{men}| \ll \chi$ and $\left|\frac{h_\infty}{h^*} - 1\right| \ll 1$. Recasting $h_\infty$ in terms of $h^*$ as

$$h_\infty = (1-\varepsilon)h^* \quad (28)$$

$|\varepsilon| \ll 1$ and substituting Eq. (28) into Eq. (26) we obtain an equation for $\varepsilon$

$$1 - \frac{2\alpha_l(m-n)}{(m-1)(n-1)} = \alpha_l(m-n)\varepsilon \frac{r}{h^*}. \quad (29)$$

Deriving Eq. (29) from Eq. (26), we, in Eq. (26), have: (a) dropped the first, the second, the seventh and the eight terms; (b) dropped $h_\infty/h^*$ in the third term and in the round brackets of the fourth term; (c) put $h_\infty = h^*$ in the fifth and the sixth terms; and (d) used $\left(\frac{1}{1-\varepsilon}\right)^m - \left(\frac{1}{1-\varepsilon}\right)^n = (m-n)\varepsilon$ in the {} brackets of the fourth term. Substituting Eq. (28) also into Eq. (22) and taking into account that $r \gg h_\infty$ and $|\varepsilon| \ll 1$, we obtain

$$P_{men} = -\frac{\gamma}{r}\left(1 + \alpha_l\,(m-n)\varepsilon\frac{r}{h^*}\right). \tag{30}$$

Substituting $\alpha\,(m-n)\,\varepsilon\,r/h^*$ from Eq. (29) into Eq. (30) and then using Eq. (23) we obtain

$$P_{men} = -\frac{2\gamma}{r}\left(1 - \frac{\alpha_l\,(m-n)}{(m-1)(n-1)}\right) \quad \text{and} \quad r_{men} = \frac{r}{1-\frac{\alpha_l\,(m-n)}{(m-1)(n-1)}}. \tag{31}$$

As in Section II, we determine the equilibrium contact angle $\theta_e$ as tangent of the angle at which the circle $r_{men}$ crosses the capillary wall (Fig. 3),

$$\tan(\theta_e) = \frac{(r_{men}^2 - r^2)^{1/2}}{r}. \tag{32}$$

Substituting $r_{men}$ from Eq. (31) into Eq. (32), we obtain

$$\tan(\theta_e) = (\alpha_l)^{1/2}\frac{\left(\frac{2\,(m-n)}{(m-1)\,(n-1)} - \alpha_l\left(\frac{(m-n)}{(m-1)\,(n-1)}\right)^2\right)^{1/2}}{1 - \alpha_l\frac{(m-n)}{(m-1)(n-1)}}. \tag{33}$$

Thus, we demonstrate that $\theta_e$ is independent of the capillary radius and matches the equilibrium contact angle for large droplets obtained in [1] as well as in the case of large capillary slabs, Eq. (19).

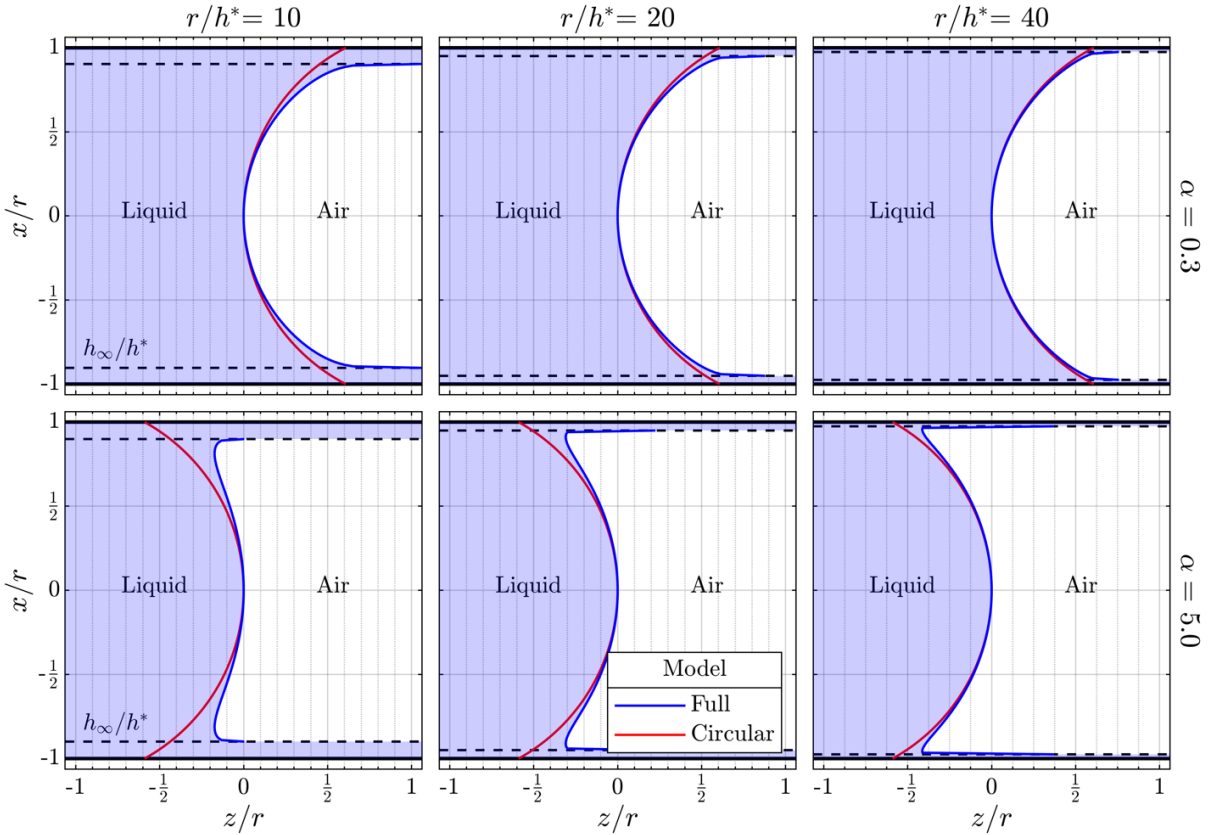

FIG. 4. Meniscus profiles calculated for various values $r/h^*$ and $\alpha$ using the full model and a circular meniscus model; $\alpha = 0.3$ corresponds to $\theta_e = 27.4°$ and $\alpha_l = 5.0$ corresponds to $\theta_e = 151.0°$.

In Fig. 4, we show the meniscus profiles calculated from Eqs. (24) – (26), the full model, against the meniscus profiles calculated by spherical meniscus model, Eq. (31). In these simulations, we, as in Section 2, have used the set of inter-molecular force exponents $m = 9$ and $n = 3$ [5], and $\alpha = 0.3$ and $\alpha = 5$ which corresponds to $\theta_e = 27.4°$ and $\theta_e = 151.0°$ respectively for different ratios of $r$ to $h^*$. As expected, that except for aforementioned foot feature (Fig. 4), both models predict very similar meniscus shapes down to $\frac{r}{h^*} > 20$.

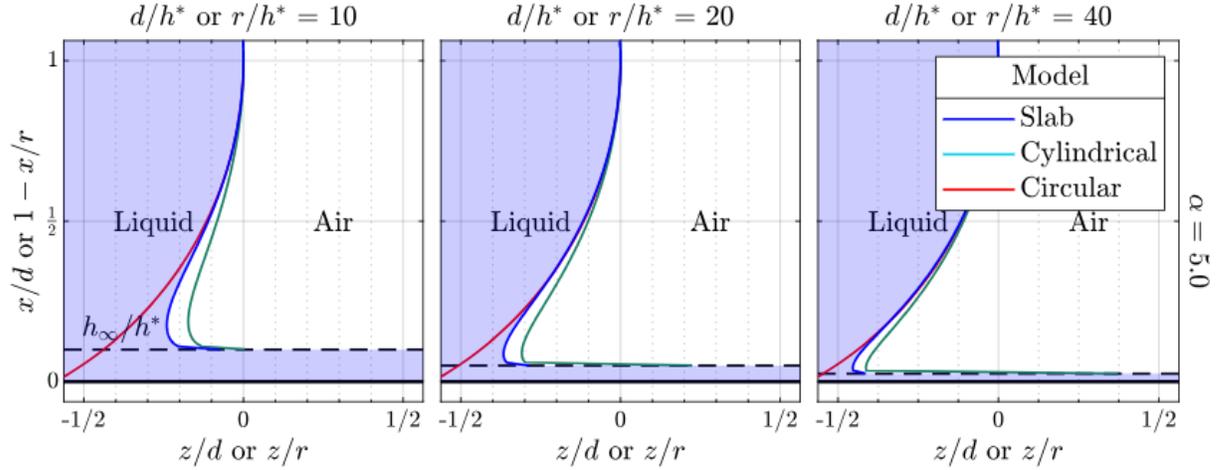

FIG. 5. Comparison of the meniscus profiles in slab and cylindrical capillaries calculated for various values $d/h^*$ and $r/h^*$, and different $\alpha$ using the full models and a circular meniscus model; $\alpha = 0.3$ corresponds to $\theta_e = 27.4°$ and $\alpha_l = 5.0$ corresponds to $\theta_e = 151.0°$.

In Fig. 5, we compare the meniscus profiles in the slab and cylindrical capillaries calculated by the full models and the circular models. As it was expected, far from the wall where the disjoining pressure is small, the differences in the "full" meniscus profiles decrease with an increase in the gap/radius of the capillaries; the circular models are the same for both geometries. However, at the walls where disjoining pressure is strong, the differences between the meniscus profiles in cylindrical and slab capillaries are different even when $d = r \gg h^*$. This is because pressures in the cylindrical and the slab capillaries are different, Eqs. (4) and (23), and the additional surface tension term in cylindrical capillaries is absent in slab case, Eqs. (1) and (20).

**IV. Dewetting of wet capillary wall**

In a capillary, a set of equation describing the motion of the meniscus can be written as,

$$\rho \frac{\partial \boldsymbol{v}}{\partial t} + \rho(\boldsymbol{v} \cdot \nabla)\boldsymbol{v} = -\nabla p + \mu \nabla \cdot (\nabla \mathbf{v}) + \kappa + \left(\gamma \kappa + \chi \left\{\left(\frac{h^*}{h}\right)^m - \left(\frac{h^*}{h}\right)^n\right\}\right)\delta_s \boldsymbol{n}, \qquad (34)$$

$$(\nabla \cdot \mathbf{v}) = 0, \qquad (35)$$

where Eq. (34) is the incompressible Navier-Stokes equation, and Eq. (35) is volume conservation equation; $\rho$ and $\mu$ are the mass density and the viscosity of the liquid, respectively, $p$ is the pressure; $\boldsymbol{v}$ is the velocity vector; $\delta_s$ is the $\delta$- function describing the position of the free surface (meniscus); $\kappa$ is the curvature of the meniscus; and $\chi\left\{\left(\frac{h^*}{h}\right)^m - \left(\frac{h^*}{h}\right)^n\right\}$ is the disjoining pressure in which $h$ is the distance between the meniscus to the capillary wall, Fig. 3; and $\boldsymbol{n}$ is a normal vector to the meniscus. This set of equations was used in [5] to describe thin film rapture using Gerris, an open-source FD solver. In our research we solve Eqs. (34) and (35) with the non-slip boundary conditions at the capillary wall using Basilisk, another open-source CFD software, in which we include the disjoining pressure similarly as it was done in [5]. It should stress that in steady state, Eqs. (34) and (35) are equivalent to Eq. (1) and Eq. (20) correspondingly for slab and cylindrical capillaries.

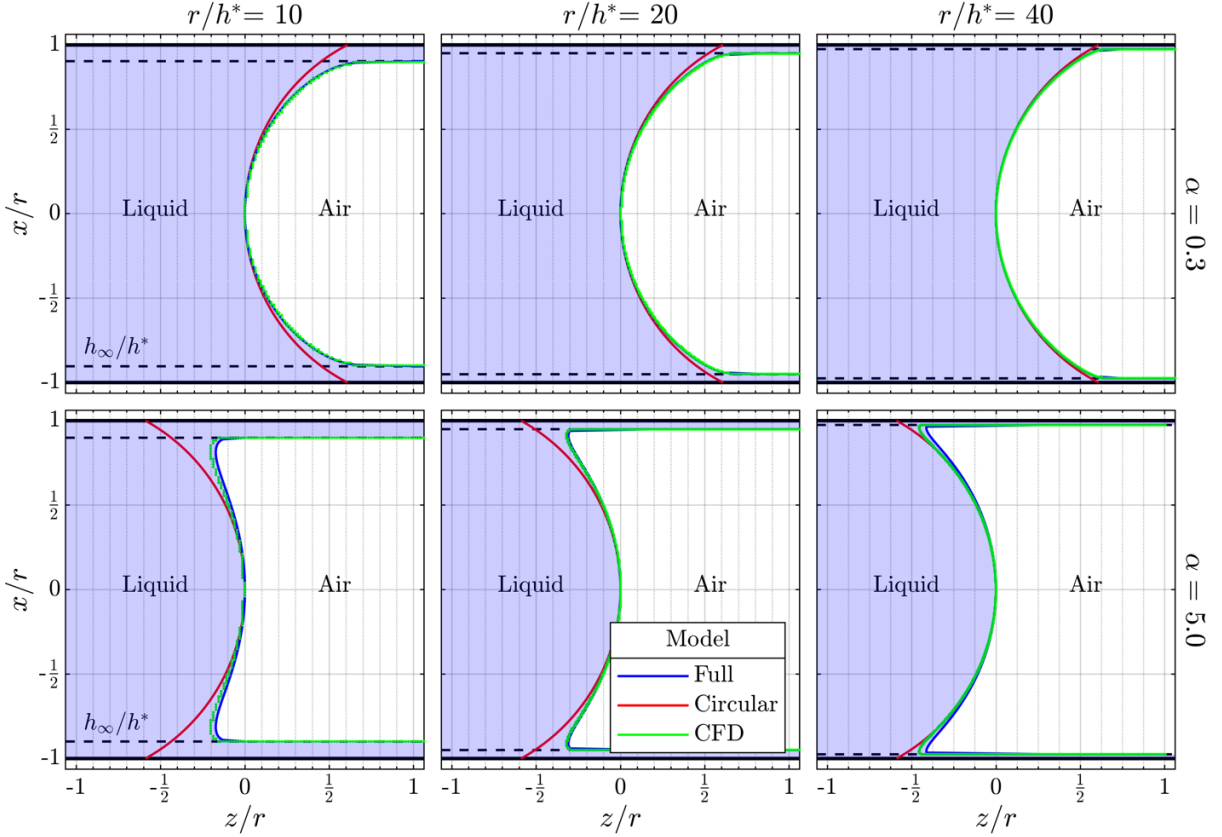

FIG. 6. Comparison of the meniscus profiles in cylindrical capillaries calculated for various values and $r/h^*$, and different $\alpha$ using the full model, the circular meniscus model, and the CFD modeling (shown the cross-sections the of the liquid-air interface with the cells); $\alpha = 0.3$ corresponds to $\theta_e = 27.4°$ and $\alpha_l = 5.0$ corresponds to $\theta_e = 151.0°$.

In Figs. 6 we show the shapes of steady state meniscus calculated by Basilisk against the steady state solution of Eq. (20) with $\alpha = 0.3$ and 5.0 for different $r/h^*$ and, in Fig. 7, show the dynamic of the meniscus motion for $\alpha = 7.0$ and 8.0. In our Basilisk simulations, we use the following set of parameters: fluid-1 -- $\rho_1 = 1050$ and $\mu_1 = 1.5$; fluid-2 -- $\rho_2 = 1.0$ and $\mu_2 = 0.018$; $\gamma = 0.05$; the equilibrium height of the film was set to $h^* = 1$, while $\chi$ was adjusted to control $\alpha$, Eq. (9). We specifically used $\chi = 0.015$, 0.25, 0.35, and 0.4 for $\alpha = 0.3, 5.0, 7.0,$ and 8.0 respectively. The initial shape of the meniscus in all our simulation was set as "flat" surface perpendicular to the capillary wall.

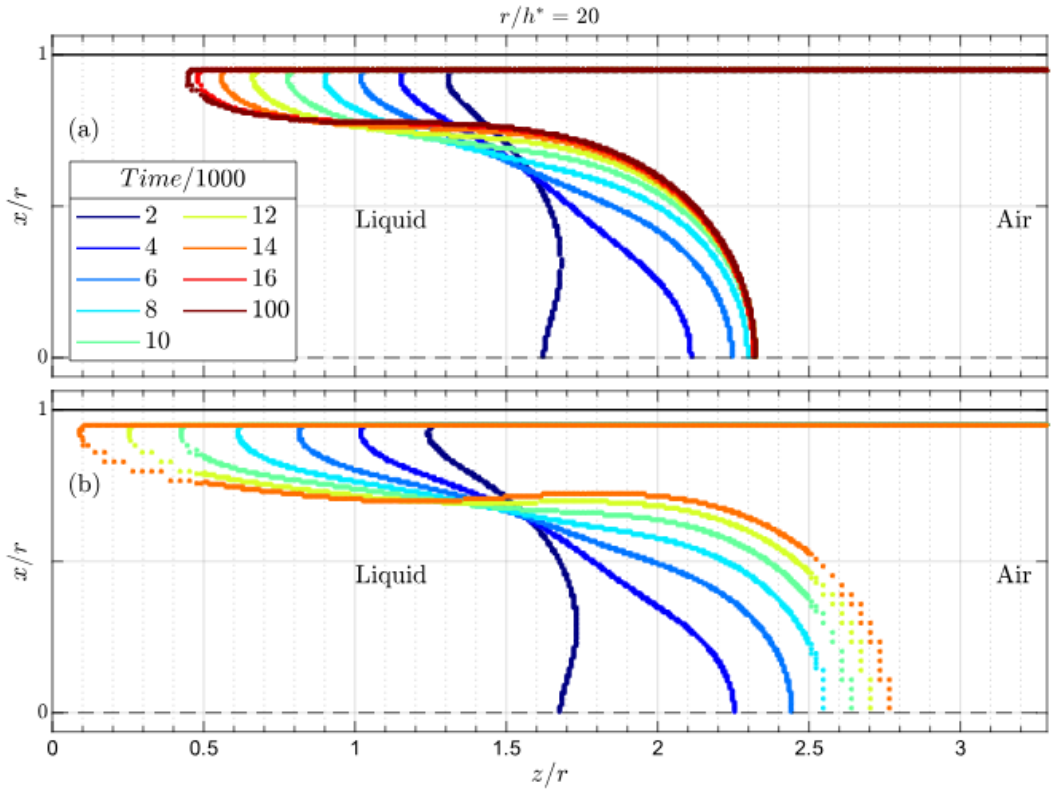

FIG. 7. Dynamics of the meniscus motion vs. time in a cylindrical capillary; shown the cross-sections of the liquid-air interface with the cells for $\alpha = 7$ (Fig. a) and $\alpha = 8$ (Fig. b). In case a, the meniscus settles to a steady state, and case b corresponds to dewetting.

As one can see from Fig. 6, the CFD simulations agree with the solution of Eq. (20) (see the previous paragraph). In Fig. 7a, the case of $\alpha = 7.0$, the motion of the meniscus settles to steady state at time about $10^6$. However, in case of $\alpha = 8.0$, Fig. 7b, the meniscus continues developing not settling to a steady state. This situation corresponds to dewetting where the liquid detaches from the wall leaving at the wall a non-removable thin liquid film. It should be noted that substituting $\alpha = 7.0$ into Eq. (33) gives imaginary value, although our Basilisk simulation predicts the steady state, Fig. 7a. The reason for this is that Eq. (33) is an asymptotic solution of Eq. (20) where the radius of the capillary is much larger than the equilibrium thickness of the film; in Fig. 7, this ratio was 20.

Finally, we checked the conversion of the obtained numerical results using different levels in our Basilisk simulation; at level 8 and 9 the numerical results were almost identical for all cases.

## V. Concluding remarks

In this article, we have demonstrated that a novel formula for the equilibrium contact angle [1] obtained in the case of a droplet placed on a wetted substrate covered by precursor non-removable thin liquid films is applicable to the case of a wetted capillaries as well. As in the case of a droplet at the wetted substrate [1], the shape of the static meniscus is obtained by balancing the liquid surface tension force against the adhesion force of the fluid to the substrate as described in the framework of disjoining pressure model. We showed that in the limit of large capillaries, i.e., $r \gg h^*$ or $d \gg h^*$ it is possible to define a contact angle that was coincident with equilibrium contact angle obtained in [1]. For both small and large contact angles, we found excellent agreement between the classical circular-cap meniscus profiles and our models for capillaries as small as $d, r \geq 20h^*$.

We also showed that when the intermolecular forces between the solid and liquid molecules became much larger than surface tension, in the framework of disjoining pressure model where $\frac{\chi h^*}{\gamma} > \frac{2(m-1)(n-1)}{m-n}$ and the radius of a cylindrical capillary or the gap of a slab capillary are much larger than the equilibrium thickness of the non-removable thin liquid film $h^*$, the meniscus in the capillary is not stable, the equilibrium contact angle does not exist, and the liquid detached from the wall leaving the non-removable thin liquid film. The dynamics of dewetting of the capillary wall was studied using Basilisk, an open CFD software, in which we include the disjoining pressure model.

Our research implies that the formula for the equilibrium contact angle suggested in [1] is, apparently, universal regardless of the geometry of capillary and the form of droplets when the characteristic dimensions of droplets or capillaries are much larger than the thickness of the non-removable, thin liquid film covering the substrates.

**ACKNOWLEDGEMENTS**

The authors would like to express their sincere gratitude to Dan Barnett for his support of this project, Sean Cahill for his valuable remarks, and Matthew Aubrey for helpful discussions.

**AUTHORS DECLARATIONS**

**Conflict of interest**

Authors have no conflict to disclose

**DATA AVAILABILITY**

The data that support the findings of this study are available from the corresponding author upon reasonable request.

**APPENDIX A: Derivation OF EQS. (24) and (25)**

We proceed by multiplying Eq. (21) by $h\frac{dh}{dz}$ that allows presenting this equation in a divergent form

$$\frac{d}{dz}\left(\gamma\frac{h}{\left(1+\left(\frac{dh}{dz}\right)^2\right)^{0.5}} - \left(\gamma\frac{1}{r-h_\infty} + \chi\left\{\left(\frac{h^*}{h_\infty}\right)^m - \left(\frac{h^*}{h_\infty}\right)^n\right\}\right)\frac{h^2}{2} + \frac{\chi r h^*}{m-1}\left(\frac{h^*}{r-h}\right)^{m-1} - \frac{\chi r h^*}{n-1}\left(\frac{h^*}{r-h}\right)^{n-1} - \frac{\chi(h^*)^2}{m-2}\left(\frac{h^*}{r-h}\right)^{m-2} + \frac{\chi(h^*)^2}{n-2}\left(\frac{h^*}{r-h}\right)^{n-2}\right) = 0. \qquad (A1)$$

Integrating Eq. (A1) we obtain

$$\gamma\frac{h}{\left(1+\left(\frac{dh}{dz}\right)^2\right)^{0.5}} - \left(\gamma\frac{1}{r-h_\infty} + \chi\left\{\left(\frac{h^*}{h_\infty}\right)^m - \left(\frac{h^*}{h_\infty}\right)^n\right\}\right)\frac{h^2}{2} +$$

$$\frac{\chi r h^*}{m-1}\left(\frac{h^*}{r-h}\right)^{m-1} - \frac{\chi r h^*}{n-1}\left(\frac{h^*}{r-h}\right)^{n-1} - \frac{\chi(h^*)^2}{m-2}\left(\frac{h^*}{r-h}\right)^{m-2} + \frac{\chi(h^*)^2}{n-2}\left(\frac{h^*}{r-h}\right)^{n-2} = C \qquad (A2)$$

where $C$ is a constant of integration. Application of boundary conditions, $\frac{dh}{dz}\underset{z\to\infty}{\to} 0$ and $h\underset{z\to\infty}{\to} r - h_\infty$ (Fig. 3), gives

$$C = \gamma\frac{r-h_\infty}{2} - \left(\chi\left\{\left(\frac{h^*}{h_\infty}\right)^m - \left(\frac{h^*}{h_\infty}\right)^n\right\}\right)\frac{(r-h_\infty)^2}{2} +$$

$$\frac{\chi r h^*}{m-1}\left(\frac{h^*}{h_\infty}\right)^{m-1} - \frac{\chi r h^*}{n-1}\left(\frac{h^*}{h_\infty}\right)^{n-1} - \frac{\chi(h^*)^2}{m-2}\left(\frac{h^*}{h_\infty}\right)^{m-2} + \frac{\chi(h^*)^2}{n-2}\left(\frac{h^*}{h_\infty}\right)^{n-2}. \qquad (B3)$$

Substituting $C$ from Eq. (B3) into Eq. (B2) we obtain

$$\gamma \frac{h}{\left(1+\left(\frac{dh}{dz}\right)^2\right)^{0.5}} = \left(\gamma \frac{1}{r-h_\infty} + \chi\left\{\left(\frac{h^*}{h_\infty}\right)^m - \left(\frac{h^*}{h_\infty}\right)^n\right\}\right)\frac{h^2}{2} -$$

$$\frac{\chi r h^*}{m-1}\left(\frac{h^*}{r-h}\right)^{m-1} + \frac{\chi r h^*}{n-1}\left(\frac{h^*}{r-h}\right)^{n-1} + \frac{\chi(h^*)^2}{m-2}\left(\frac{h^*}{r-h}\right)^{m-2} - \frac{\chi(h^*)^2}{n-2}\left(\frac{h^*}{r-h}\right)^{n-2} +$$

$$\frac{\gamma(r-h_\infty)}{2} - \chi\left\{\left(\frac{h^*}{h_\infty}\right)^m - \left(\frac{h^*}{h_\infty}\right)^n\right\}\frac{(r-h_\infty)^2}{2} +$$

$$\frac{\chi r h^*}{m-1}\left(\frac{h^*}{h_\infty}\right)^{m-1} - \frac{\chi r h^*}{n-1}\left(\frac{h^*}{h_\infty}\right)^{n-1} - \frac{\chi(h^*)^2}{m-2}\left(\frac{h^*}{h_\infty}\right)^{m-2} + \frac{\chi(h^*)^2}{n-2}\left(\frac{h^*}{h_\infty}\right)^{n-2}. \quad \text{(B4)}$$

Dividing Eq. (B4) by $h^* \gamma$, this equation reduces to the following form

$$\frac{1}{\left(1+\left(\frac{dh}{dz}\right)^2\right)^{0.5}} = B_c, \quad \text{(B5)}$$

$$B_c(h) = \frac{1}{2}\left(\frac{h^*}{r-h_\infty} + \alpha\left\{\left(\frac{h^*}{h_\infty}\right)^m - \left(\frac{h^*}{h_\infty}\right)^n\right\}\right)\frac{h}{h^*} +$$

$$\alpha\frac{h^*}{h}\begin{pmatrix} -\frac{r}{h^*}\frac{1}{m-1}\left(\frac{h^*}{r-h}\right)^{m-1} + \frac{r}{h^*}\frac{1}{n-1}\left(\frac{h^*}{r-h}\right)^{n-1} + \frac{1}{m-2}\left(\frac{h^*}{r-h}\right)^{m-2} - \frac{1}{n-2}\left(\frac{h^*}{r-h}\right)^{n-2} + \\ \frac{1}{2\alpha}\frac{r-h_\infty}{h^*} - \frac{1}{2}\left\{\left(\frac{h^*}{h_\infty}\right)^m - \left(\frac{h^*}{h_\infty}\right)^n\right\}\left(\frac{r-h_\infty}{h^*}\right)^2 + \\ \frac{r}{h^*}\frac{1}{m-1}\left(\frac{h^*}{h_\infty}\right)^{m-1} - \frac{r}{h^*}\frac{1}{n-1}\left(\frac{h^*}{h_\infty}\right)^{n-1} - \frac{1}{m-2}\left(\frac{h^*}{h_\infty}\right)^{m-2} + \frac{1}{n-2}\left(\frac{h^*}{h_\infty}\right)^{n-2} \end{pmatrix}, \quad \text{(B6)}$$

where index "c" corresponds to cylindrical geometry and $\alpha$ is given by Eq. (9). Solving Eq. (B5) for $\frac{dh}{dz}$ we obtain Eq. (24), and Eq. (B6) corresponds to Eq. (25).

**APPENDIX B: DERIVATIO OF EQ. (27)**

The second term in the RHS of Eq. (25) is

$$A = \alpha\frac{h^*}{h}\begin{pmatrix} -\frac{r}{h^*}\frac{1}{m-1}\left(\frac{h^*}{r-h}\right)^{m-1} + \frac{r}{h^*}\frac{1}{n-1}\left(\frac{h^*}{r-h}\right)^{n-1} + \frac{1}{m-2}\left(\frac{h^*}{r-h}\right)^{m-2} - \frac{1}{n-2}\left(\frac{h^*}{r-h}\right)^{n-2} + \\ \frac{1}{2\alpha}\frac{r-h_\infty}{h^*} - \frac{1}{2}\left\{\left(\frac{h^*}{h_\infty}\right)^m - \left(\frac{h^*}{h_\infty}\right)^n\right\}\left(\frac{r-h_\infty}{h^*}\right)^2 + \\ \frac{r}{h^*}\frac{1}{m-1}\left(\frac{h^*}{h_\infty}\right)^{m-1} - \frac{r}{h^*}\frac{1}{n-1}\left(\frac{h^*}{h_\infty}\right)^{n-1} - \frac{1}{m-2}\left(\frac{h^*}{h_\infty}\right)^{m-2} + \frac{1}{n-2}\left(\frac{h^*}{h_\infty}\right)^{n-2} \end{pmatrix}. \quad \text{(B1)}$$

Let us analyze A-term when $h \to 0$. Using the second order of the Taylor expansion, $(r-h)^l = r^l\left(1 - l\frac{h}{r} + \frac{l(l-1)}{2}\left(\frac{h}{r}\right)^2\right)$ we obtain

$$A \approx \alpha_l \frac{h^*}{h}\left(-\frac{1}{m-1}\left(\frac{h^*}{r}\right)^{m-2}\left(1 + (m-1)\frac{h}{r} + \frac{(m-1)m}{2}\left(\frac{h}{r}\right)^2\right) + \right.$$

$$\frac{1}{n-1}\left(\frac{h^*}{r}\right)^{n-2}\left(1+(n-1)\frac{h}{r}+\frac{(n-1)n}{2}\left(\frac{h}{r}\right)^2\right)+$$

$$\frac{1}{m-2}\left(\frac{h^*}{r}\right)^{m-2}\left(1+(m-2)\frac{h}{r}+\frac{(m-2)(m-1)}{2}\left(\frac{h}{r}\right)^2\right)-$$

$$\frac{1}{n-2}\left(\frac{h^*}{r}\right)^{n-2}\left(1+(n-2)\frac{h}{r}+\frac{(n-2)(n-1)}{2}\left(\frac{h}{r}\right)^2\right)+$$

$$\frac{1}{2\alpha_l}\frac{r-h_\infty}{h^*}-\frac{1}{2}\left\{\left(\frac{h^*}{h_\infty}\right)^m-\left(\frac{h^*}{h_\infty}\right)^n\right\}\left(\frac{r-h_\infty}{h^*}\right)^2+$$

$$\frac{r}{h^*}\frac{1}{m-1}\left(\frac{h^*}{h_\infty}\right)^{m-1}-\alpha_l\frac{r}{h^*}\frac{1}{n-1}\left(\frac{h^*}{h_\infty}\right)^{n-1}-\frac{1}{m-2}\left(\frac{h^*}{h_\infty}\right)^{m-2}+\alpha_l\frac{1}{n-2}\left(\frac{h^*}{h_\infty}\right)^{n-2}\Bigg)=$$

$$=\frac{\alpha_l}{2}\left(\left(\frac{h^*}{r}\right)^n-\left(\frac{h^*}{r}\right)^m\right)\frac{h}{h^*}. \tag{B2}$$

In Eq. (B2), we have taken into account Eq. (26).